# Resize, Remix, Regen: Frankensteining IoT Design Methods

**Albrecht Kurze**

Chemnitz University of Technology
Faculty of Computer Science
Chair Media Informatics
Chemnitz, Germany
albrecht.kurze@informatik.tu-chemnitz.de

**Abstract**
There are numerous IoT design methods. Previous research shows that all of them have their strengths, but also their limitations. None of them is a universal, all-purpose method. However, experts often view these methods as more versatile than their creators intended. Therefore, analyzing existing methods and tools, as well as rearranging and combining their approaches and components - just as Frankenstein did with his creature - offers the possibility of new creations that may be better than any single method previously. We present the idea and concept of Frankensteining, which is based on the repeated application of IoT design methods in various contexts. We present a practical Frankensteining creation that was used in a workshop, our own methods, and a serial Frankensteining approach that was tested in an educational context. We conclude with a discussion on Frankensteining and invite other experts and practitioners to share their perspectives and experiences.

## 1 Introduction

*"Design methods are like toothbrushes.*
*Everyone uses them, but no one likes to use someone else's."*
~ attributed to John Zimmerman (Harrison & Tatar, 2011)

There are many design methods for the Internet of Things (IoT). Table 1 lists a non-exhaustive selection. Yet, exchange between researchers and practitioners in this field has been limited. Most methods are developed and applied more in isolation than in cooperation, with only occasional references to one another. Consequently, little is known about how a given method should really work, how it actually works, or about its strengths and limitations. A series of community-organized events has begun to change this situation. Workshops at ThingsCon conferences (2016-2019), a ThingsCon salon (2018), and academic workshops at expert level (Kurze et al., 2019) and conferences IASDR 2021 (Berger et al., 2022) and ACM C&C (Mora et al., 2022) helped to bring together creators and users of IoT design methods. Participants discovered that hands-on experience with the methods in a workshop setting is far more informative than merely reading about them in publications. Building on this momentum, we analyzed how experts evaluate the methods (Kurze et al., 2019).

Influenced by our experiences in the workshops, we and other participants had at least a vague sense that there might be a lot of potential in combining existing methods. As a result, one of the workshop participants coined the term "Frankensteining." And indeed, if you consider Mary Shelley's book Frankenstein, published in 1818, that was exactly what we had in mind and resonates very well with the ideas of "resize", "remix" and "regen":

a) Adopt certain parts and steps, omit others, scaling up or down, and adapt parts or complete methods. Add new methods only as needed. **> Resize**
b) Combine parts and steps from different methods into mashups, e.g. for different design phases and iterations or to incorporate different perspectives (e.g., desirability, feasibility and viability). **> Remix**
c) The process of bringing the methods and new creations (back) to life to demonstrate their usefulness and practical value, e.g. in workshops, and to reflect on and document these use cases. **> Regen**

We wanted to gain a better understanding of how such a Frankensteining of methods could actually be implemented and what benefits it offers. In the years following the aforementioned workshop, however, new topics emerged, research projects and grants came to an end, former experts left the topic, and some methods were abandoned. Nevertheless, we stuck with the idea during this time, collected additional methods, devised approaches, and gained experience in the practical application of Frankensteining methods. This article presents some of these creations and use cases to illustrate possible approaches, discusses some of the insights gained, and will hopefully inspire others.

**Table 1**: Overview of selected IoT design methods with references.

| **Method** | **References (paper or article etc.) / URL (material etc.)** |
|---|---|
| Cards'n'Dice | (Berger et al., 2019; Lefeuvre et al., 2016, 2017; Kurze et al., 2016)<br>https://www.tu-chemnitz.de/informatik/mi/projekte/nebeneinander-miteinander/portfolio/loaded-dice.html (last accessed 2026/04) |
| Co-create the IoT | (Stembert, 2017; van Kranenburg et al., 2014)<br>https://stembertdesign.com/cocreation-and-the-iot.html (last accessed 2026/04) |
| Internet of Tangible Things Toolkit | (Angelini et al., 2018)<br>https://sites.google.com/view/iott-design-kit (last accessed 2026/04) |
| IoT Design Deck | (Dibitonto et al., 2017)<br>https://www.iotdesigndeck.com/ (last accessed 2026/04) |
| IoT Design Kit | (De Roeck et al., 2019)<br>https://iotdesignkit.studiodott.be/ (abandoned)<br>https://miro.com/templates/iot-design-kit/ (last accessed 2026/04) |
| KnowCards | (Aspiala, 2014)<br>https://boingboing.net/2014/12/10/cards-for-brainstorming-produc.html (last accessed 2026/04)<br>https://www.designswarm.com/know-cards/ (last accessed 2026/04) |
| Mapping the IoT | (Vitali et al., 2016; Vitali & Arquilla, 2018)<br>https://mappingtheiot.polimi.it/ (broken, last accessed 2026/04)<br>https://mappingtheiot.polimi.it/downloads/ (broken but still downloadable, last accessed 2026/04) |
| Tiles IoT Toolkit | (Mora, Asheim, et al., 2016; Mora, Divitini, et al., 2016; Mora et al., 2017)<br>https://www.tilestoolkit.io/ (last accessed 2026/04)<br>https://www.tu-chemnitz.de/informatik/mi/projekte/nebeneinander-miteinander/portfolio/tiles-iot-toolkit-de.html (Digital Edition, German, last accessed 2026/04) |

# 2 Frankensteining theory

In a previous study (Kurze et al., 2019), we examined how experts mapped design methods to stages of the Double Diamond process model (Design Council, 2005). Our findings showed that it is not easy to map entire design methods onto the Double Diamond process model. Often, the mappings were vague and spanned the entire design process (two examples in Figure 1). While the developers of a design method may have had a clear idea of how and when their method should be used, other experts often viewed the methods as more versatile and not as clearly positioned.

We have also shown that there is no single one-fits-all universal IoT design method, as the methods clearly have different strengths and focuses. Therefore, we have presented an alternative context-dependent multidimensional mapping approach inspired by Sanders & Stappers (2014) that allows other designers and researchers to identify and select methods based on identified strengths and suitability.

Our findings also provided insights into what caused the fuzziness of the mappings along the design process model. We launched a new attempt to map IoT design methods along the Double Diamond process model (Figure 2). This time, however, we used a higher level of granularity that identifies and accounts for the various steps in the methods and the components and materials involved – an approach that some of the expert mappings had already demonstrated previously (Kurze et al., 2019). We again opted for the Double Diamond model, as this allows for the inclusion of insights from previous work and enables a comparison with existing expert mappings of methods. Depending on the intended use, the same element or methods may appear multiple times in a mapping.

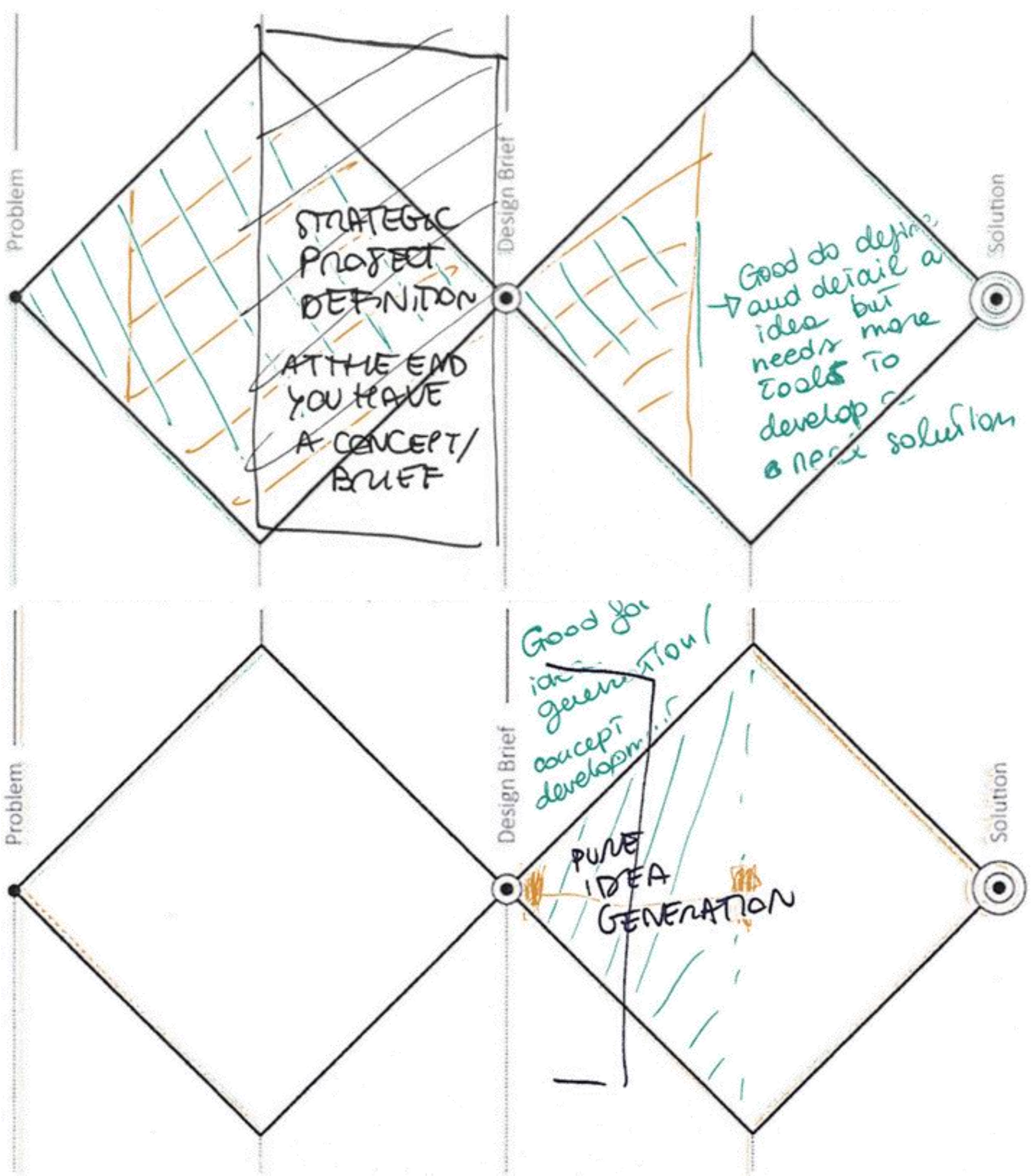

**Figure 1**: Mapping of the *IoT Design Kit* (a, top) and *Tiles IoT Inventor Toolkit* (b, bottom) on the Double Diamond design process model as overlays of three selected expert mappings each (color coded). The mapping by the creator of the according method is shown in orange.

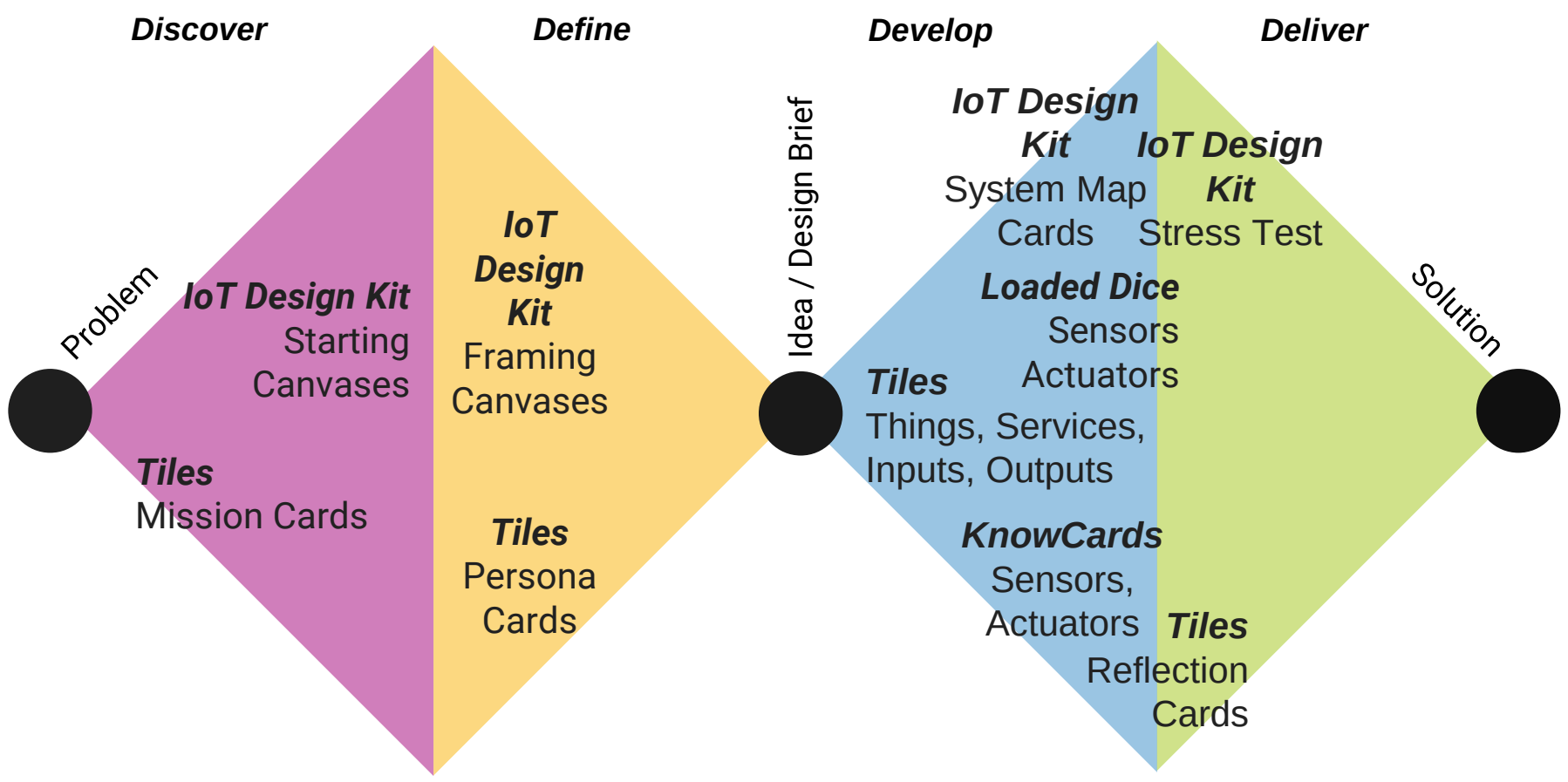

**Figure 2**: Our attempt of mapping components of selected IoT design methods for combination in a design process along the Double Diamond process model.

Such an approach can serve as a starting point for process- or material-based Frankensteining approaches. Nevertheless, it is essential to understand what a method involves, what it consists of, and how it works. This necessarily requires that one has already applied and become familiar with the methods before combining them. Otherwise, it might be difficult to implement them smoothly or achieve the expected benefits.

A resulting "Frankenstein" creation can be a mashup of materials (e.g. cards or canvases) and methodological steps in the design process that form something entirely new. Therefore, the concept differs somewhat from a simple combination of entire methods that could be applied sequentially or used interchangeably in parallel. The creation should be usable, applicable in practice, and beneficial in terms of the outcome, e.g. in workshops. Upon closer examination of the mappings created by experts for various IoT design methods (Kurze et al., 2019), some useful comments indicate that the experts had already identified the methods' strengths and weaknesses. Our analysis and our own experiences provided us with further clues regarding possible combinations.

# 3 Use case examples

The following three use case examples illustrate what a "Frankensteining" of IoT design methods, incorporating resize, remix, and regen approaches, might look like.

## 3.1 Use case I: A new nameless creation

We have launched a first hands-on workshop, in which we combined the *IoT Design Kit* and the *Tiles IoT Inventor Toolkit* to explore their potential benefits. Both tools are ideally suited to pool their respective strengths and bridge gaps in their respective methodologies during the design process through a new combination, without increasing complexity or creating too much redundancies.

The *IoT Design Kit* is primarily intended for companies that wish to integrate the Internet of Things into their business operations. The method is typically guided by its developers during sessions lasting several hours. Idea generation should be understood as part of a broader process, as the method's focus is not only on generating ideas but also on corporate strategy and positioning with regard to the IoT. The method offers various starting points and the option to carry out individual steps of the design process as standalone exercises, supported by various templates. The method deliberately provides only a few examples on cards in the various categories, e.g. for stakeholders or involved things (Table 1). The focus is on creating your own, use case specific cards using the provided blank cards. Creating a 'system map' that links people, objects, and the environment with interaction cards is a central exercise of the method.

The *Tiles IoT Inventor Toolkit* promises to help non-experts generate ideas or develop IoT products in a short amount of time. The cards come with a single all-in-one canvas and a guide containing instructions on how to use the cards (no facilitator required). The method is not designed to delve too deeply into the details of product design, as its primary focus is on developing a basic understanding and generating ideas quickly (45–60 minutes per session). The method offers far more examples for inspiration in almost all categories—such as 'Things' than the *IoT Design Kit* (Table 1). *Tiles* also includes cards for aspects and categories not covered in the *IoT Design Kit*.

Both methods share a step-by-step approach that provides clear guidance through the design process. We primarily used the steps defined by the *IoT Design Kit* based on the included canvases. We selected five out of ten canvases. For problem formulation, we used a selection of 11 of the 19 'Misc' cards and kept the *Tiles* 'Mission' cards on hand to use as needed. We combined the 'People' and 'Persona' cards to define the stakeholders. The main Frankensteining took place during the creation of a system map by adding materials as well as process steps. Figure 3 shows the various materials used in the different steps.

**Table 2**: Comparing the components of the *IoT Design Kit* and the *Tiles IoT Inventor Toolkit.*

| Component | IoT Design Kit | Tiles IoT Toolkit |
|---|---|---|
| canvas(es) | 10 in total to select from<br>~5 to use per run | 1 all-in-one in A0<br>format for all steps |
| framing/goal | 19 misc cards | 22 mission cards<br>5 scenario cards |
| context | 15 environment cards | - |
| stakeholders | 11 person cards | 10 persona cards |
| non-human actors | 8 object cards | 42 thing cards<br>27 service cards |
| interactions | 1 interaction card for human or data driven interaction as link in system map | 10 human action cards (input)<br>11 feedback cards (output) |
| additional tech | - | 9 sensor cards |
| reflection | 9 wildcards | 14 criteria cards |

The Frankensteining was carried out in these steps using three specific approaches:

1. Canvases from the IoT Design Kit for structuring and guidance of the main steps in combination with the 'Misc' cards from the IoT Design Kit and the 'Persona' cards from Tiles.
2. 'System map' cards from the IoT Design Kit for developing a hybrid product-service system and for designing the connections (interactions) between various components, such as persons and objects. We used them in combination with the Tiles 'Thing' cards.
3. Tiles cards are used in refinement steps for services, inputs, outputs, and sensors to expand and detail the system map.

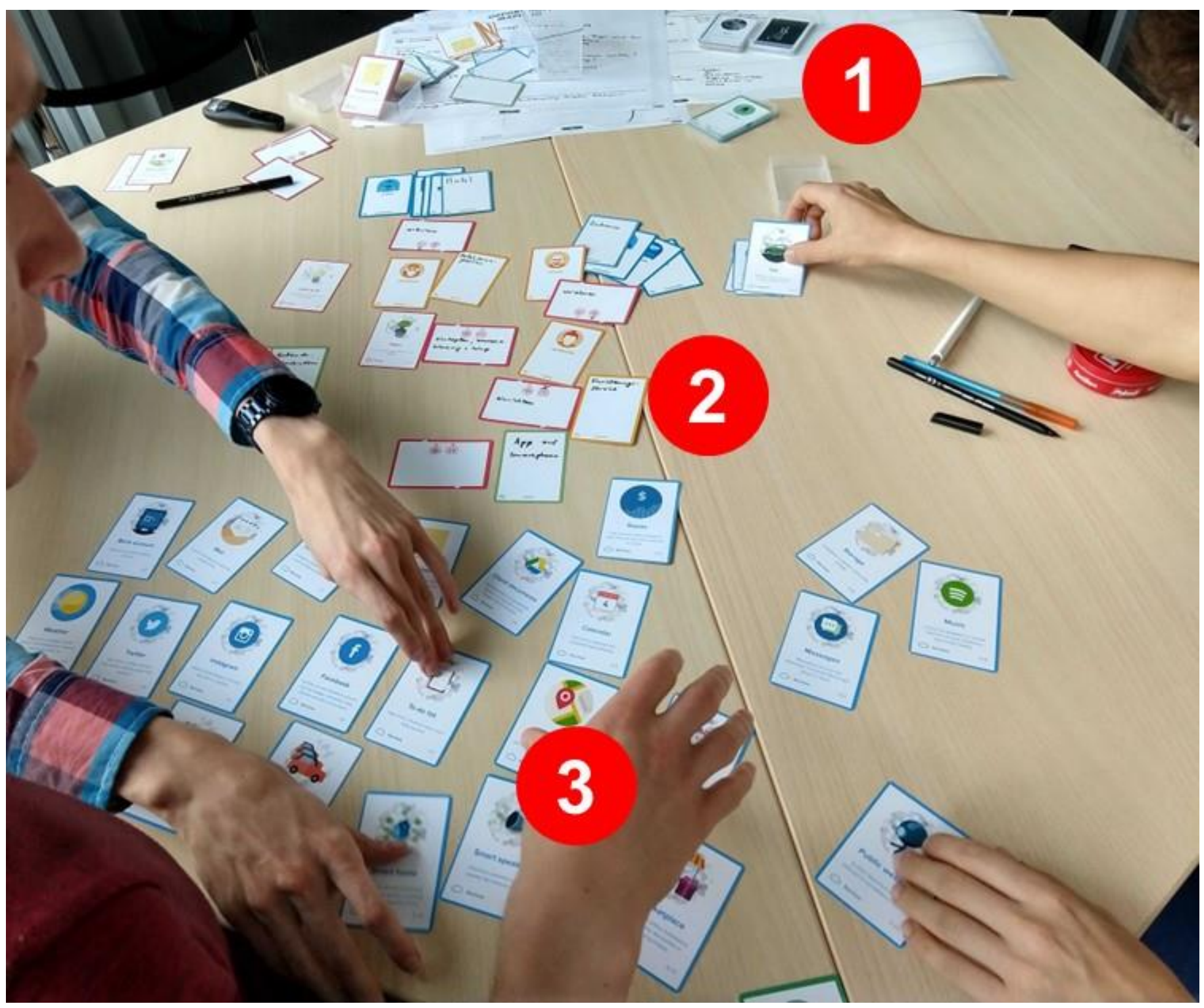

**Figure 3**: Using hands-on a Frankensteining approach with two IoT design methods in a workshop.

This Frankensteining experiment proved useful and highly successful during a workshop at our university's center for entrepreneurship. The participants had diverse professional backgrounds and varying levels of knowledge regarding design processes in general and the Internet of Things in particular. They appreciated the additional possibilities that arose from the gradual introduction of further *Tiles* cards into the *IoT Design Kit* main steps. The system maps created gained in systemic complexity compared to earlier, purely conceptual uses of the standalone *IoT Design Kit* and addressed HCI-related aspects in great detail. The combination of both methods made it possible to move from a more conceptual level to a closer to real-world implementation with much greater detail. The detailed understanding of possible system outcomes also allowed better reflection on possible implications, e.g. in terms of privacy.

## 3.2 Use case II: The Cards'n'Dice hybrid

Frankensteining can also refer to the use of real (functional) IoT technology in the IoT design process to bring ideas and concepts to 'life', similar to what Frankenstein did with the help of electricity, e.g. by replacing solely on analog representations of sensors on 'sensor' cards with actual and functional sensor components.

The *Loaded Dice* (Kurze et al., 2016; Lefeuvre et al., 2016) are such fully functional IoT design tools equipped with six real sensors and six real actuators for inputs and outputs. They can bring new dimensions of ideation, exploration, and evaluation, and open up entirely new perspectives, e.g., for idiosyncratic (Berger et al., 2019) and synesthetic (Kurze, 2022c) exploration and ideation through simple and rapid experimentation beyond the conceptual level. The use of such a tool in combination with other IoT methods, which are often only available in paper-based form as maps and canvases, can lead to interesting approaches. Particularly promising is the combination with all methods that involve sensing and actuating data, which is otherwise only limited to printed descriptions on cards.

One example of a remix is using *Loaded Dice* in conjunction with the *Tiles IoT Toolkit*. The materials and steps provided by *Tiles* can be used up to system design or the reflection step. At this point, *Loaded Dice* can be used to quickly try-out new or evaluate previously designed interactions using selected sensors and actuators, to get a sense of 'how it feels' when the system is actually implemented, and to subsequently adapt the scenario and find more suitable inputs and outputs.

Our own "Cards'n'Dice" method, which has been used in numerous workshops with various user groups (Berger et al., 2019), is in fact already a kind of hybrid that has reimagined and recombined elements of other methods. Specifically, it consists of the following components:

a) a set of cards for structured, step-by-step scenario creation using context cards for goals and actors, as well as spatial context cards inspired by other methods,
b) an adapted interaction vocabulary that builds on earlier work by Diefenbach et al. (2013) by assigning them to cards with an interaction adjective on one side and its antonym on the other (e.g., "slow" and "fast"), as well as the introduction of a new category of emotional interaction qualities (Kurze, 2022a) and new pairs of characteristics (e.g., "harsh" and "tender"),
c) the Loaded Dice themselves as fully functional design tools.

We have built on this approach and successfully extended the *Loaded Dice* concept along with accompanying cards for designing workshops to the *Wheel of Plush* (Sontopski et al., 2024), a tool for exploring for the design space of smart textiles, such as plush toys (Kurze, Chuang, et al., 2023). To this end, we have adapted the components as follows: a) The card set for the workshop scenario has been expanded to represent social relationships between the participants (e.g. between children and parents), b) the interaction vocabulary has been adapted to focus primarily on tangible interactions and tactile experiences, and c) the design tool itself has been adapted to offer even more sensors (eight) and actuators (eight) covered in plush for exploration in even more diverse and flexible configurations and combinations (Stephan et al., 2024), as well as new combined soft-sensor-actuator inflatables (Stephan et al., 2025).

### 3.4 Use case III: Serial Frankensteining in education

The third use case focuses on a university course on human-computer interaction. This advanced course emphasizes interactions beyond screens, human-data interaction, interactions with smart devices and environments, as well as human-centered design methods such as ideation and similar approaches.

For the part on human-data interaction, the course builds on the previous course concepts *Data-I* (Kurze, Köpferl, et al., 2023) as well as on real-world IoT data experiences that were collected using real IoT sensors (Kurze, 2022b) and made accessible using our *Sensorkit* (Kurze, Reuter, et al., 2025). For ethical reflection on designed sensor data collections, visualizations and very likely unintended yet wicked implications (Kurze & Bischof, 2021) we adopted the card-based serious game *Moral Agent* (Gispen, 2017). We transformed the approach in a structured canvas that guides the students in reflection on different values, e.g. to let them evaluate gains on efficiency, comfort and security etc. on the hand side and impacts on privacy, sovereignty and trust etc. on the other hand side.

At the core of the part on interaction with smart objects is the sequential application of three methods:

1. *Tiles IoT Toolkit* for initial brainstorming, scenario creation, designing systems and interactions, and a first reflection. We created and use a localized German Digital Edition of the material on a Miro board since it easily scales in larger courses.
2. *Cards'n'Dice* for exploring real-world sensor and actuator possibilities and for evaluating the designed interactions. The availability and number of *Loaded Dice* pairs is clearly a limiting factor here. We have redesigned and created a more cost-effective version that is also easier to manufacture.
3. *Internet of Tangible Things Toolk*it for mapping and linking the designed tangible interactions to the digital realm of data and connections using the original cards on a mapping canvas. We recreated the mapping canvas because, while it was described in the original publication, it was not distributed along with the card set.

In principle, this involves a mashup of various components from different methods, which are recombined and applied step by step within the framework of a larger meta-method. It is only through the combination of all three methods, where the result of one serves as the starting point for the next, that a serial (in a sense iterative) overarching method is formed to achieve the desired outcome (final result and learning objectives). In this sense, the use case is an example of how a sequential combination of powerful methods with strong elements throughout the HCD process with some adjustments, additions, and iteration in overlapping areas can be understood as a form of remix for the (re)generation of a meta-method.

The ability to understand and apply suitable methods not only for ideation but also for reflecting on implications is a key competency for the next generation of future designers and developers of IoT and smart technologies (Kurze, Bischof, et al., 2025). In this sense, Frankensteining is also a valuable approach and contribution.

## 4 Discussion

The discussion is structured around a few thought-provoking questions, which are deliberately phrased in a somewhat provocative manner and are intended to encourage us to think beyond the status quo, a point that was succinctly captured by the opening quote: *"Design methods are like toothbrushes."*

### 4.1 Create>Recreate or Recreate>Create?

*Do you have to develop your own method first before you can appreciate, adapt, recombine, or redesign existing methods?*

**Yes and no.**

Even though "yes" might sound a bit counterintuitive, there could be some truth to it. Developing a method or a tool requires a deep reflection on the elements and steps involved (as well as on what has been omitted) than simply applying a method according to instructions. Therefore, it is not only the methods and tools themselves that are necessary, but also shared knowledge. This includes insights into the design decisions made during their development, as well as well-documented use cases: what has worked well and what offers room for improvement. On the part of the users of these methods and tools, the ability is required to thoroughly analyze and understand them, as well as to reflect on one's own use and their applicability for one's own purposes. However, the Frankensteining of methods can also be taught. This includes the ability to select methods, analyze them, reflect on results, strengths, and weaknesses, and finally adapt, recombine, and redesign them.

## 4.2 Abomination or evolution?

*Is every possible Frankensteining both meaningful and feasible?*

**Probably not all of them, but some are.**

When you think of Frankenstein's creation, terms like "monster" or "abomination" might first come to mind. Different appearances of material obtained through various methods may make a creation seem a bit ugly; some steps may appear clumsy, at least on the first try. This is where "regen" comes into play; it is needed to evaluate approaches and results for their practicality and then decide how to proceed, discarding the bad ones and keeping the good ones for the next iteration. That is evolution. That is the process everyone must learn.

However, since the design methods are often somehow vague and fuzzy, it is difficult to objectively evaluate or even compare the results of different methods, including Frankensteining methods. In our experience, every use of a method, e.g. in a workshop or design process in general, is idiosyncratic, not only due to the methods and tools themselves, but also to the participants, workshop and design objectives, and facilitation. This is an aspect reflected in Frankensteining aiming to adapt to this variety, yet one that defies strict comparison. To achieve reliable results regarding the evaluation or comparison of methods or outcomes, (much) more data from uses of the methods, aimed at evaluation and comparison, would be required. This would not be truly feasible in practice, if it were even possible at all.

## 4.3 Paper or repository?

*How should we publish methods or make components available?*

**FAIR and, ideally, for free.**

The answer applies the FAIR principles for data (Wilkinson et al., 2016) to methods as well as associated tools and materials, which should be findable, accessible, interoperable, and reusable.

The availability and openness of methods and tools play a crucial role in their use – whether for their intended purpose or for Frankensteining combinations. And that is a real problem. In the academic world, a scientific paper is the medium of choice for publication. This makes knowledge about a method and a tool accessible, often along with insights into design decisions, findings on how well it worked for a specific user group, as well as a discussion of strengths and limitations. A paper also makes this knowledge permanently available and discoverable, ideally archived with a Document Object Identifier (DOI). However, this type of dissemination often lacks the availability of the tool or accompanying materials for the method, at least over a longer period or in a permanent manner. A private or institutional website is suboptimal. These are often abandoned or redesigned after only a few years, causing materials along with documentation and valuable insights to disappear. A significant number of the methods in our collection have already met this fate. Therefore, a public repository (e.g. GitHub or Zenodo) is likely the better approach. It ensures that the material remains findable and accessible, enables the organization of source files and the release of versions, records changes, and is fundamentally based on the understanding that things change and evolve, driven by a potentially large number of contributors. However, in a 'repository only approach' a brief README.md will likely lack the depth of knowledge offered by a previously mentioned paper. Nevertheless, any form of additional knowledge beyond the raw materials will help make methods and tools usable in such a way that they become (inter)operable even for others besides their creators.

When it comes to licensing, clarity and transparency play a key role. Clarity in the sense that there is a license, and openness regarding what this license permits. Here, a liberal Creative Commons (CC) license without the "non-commercial" (NC) or "no derivatives" (ND) restrictions seems to be the most suitable for reuse. We have benefited greatly from the fact that most of the methods we actually use are available for free download. With thicker paper, a standard printer, a hole punch, and a little patience and care, we were able to create copies and variations that look quite presentable thanks to their uniform card format - even for Frankensteining. Nevertheless, developing a method involves a tremendous amount of effort. Therefore, marketing a tool as a commercial product is a sensible option. This approach also helps make a tool publicly available, often through print-on-demand and similar channels. The *KnowCards* and the *IoT Design Deck* are examples of this. And replicating a method or toolkit, say with some dozen cards, isn't free either in terms of printing costs or time. The *Tiles IoT Toolkit* offers an interesting hybrid approach consisting of a free downloadable version and a ready-to-use "workshop bag" for sale.

Overall, a well-maintained repository for sharing materials and instructions, combined with a paper offering detailed insights and reflections, and a liberal CC license will serve as the best starting point for methods to enable a Resize>Remix>Regen Frankensteining approach.

## 4.4 Future work

*How should we proceed with "Frankensteining"?*

**Be curious, try out some strange experiments, and share your experiences.**

The Frankensteining approach is not limited to the IoT design methods used in the examples. We initially mapped suitable components for the *IoT Design Kit*, the *Tiles IoT Inventor Kit*, and *KnowCards*, as these could complement each other easily. However, there are other combinations that have not yet been tested. Some of these approaches may focus less on ideation and more on the implications – such as privacy or ethical aspects of the IoT and smart devices – and integrate even more than the methods mentioned above (Kurze & Berger, 2022). We are particularly interested in how functional IoT design methods and tools, such as the *Loaded Dice*, could open up new dimensions of Frankensteining. They enrich the design process beyond the printed cards with functional elements that could also offer interesting possibilities for scaling and recombination with other methods. This also includes aspects of data as design material in the human-centered design process (Gomez Ortega et al., 2023) as well as the integration of Artificial Intelligence for handling the data and opening new interaction possibilities.

The Frankensteining approach is by no means limited to design methods in the field of the Internet of Things. The described Frankensteining of *Moral Agent* is such an example in the realm of ethics methods and tool. In our research project *Simplications* (Kurze, Bischof, et al., 2025), which examines the unintended implications of simple sensor data in the smart home, we have expanded this approach beyond pure design methods. We have also tested it with game-oriented methods that include similar materials (cards and a game board), steps, rounds, or other forms of methodological guidance, and often explicit rules as well. We have adapted the *Privacy Awareness Cards* (Lorenz et al., 2024) a serious game featuring a game board and categorized cards, with an extension that introduces privacy-relevant aspects of a smart home context, such as sensor data, as well as specific spatial and social contextual features. Resizing and remixing are achieved by adding and removing specific cards from the original card set, while retaining the established game board and the steps that structure the game.

Overall, we are interested in the perspectives and experiences of other experts who use various (IoT) design methods, and perhaps even in how they have previously employed a Frankensteining-oriented approach. Therefore, we invite other experts to participate in our approach and share their views and findings.

## Acknowledgements

This research is partially funded by the German Ministry of Research, Technology and Space (BMFTR) grants FKZ 16SV9117 and FKZ 16KIS1868K. A generative AI tool (DeepL) was used for the stylistic revision and formulation of individual text passages; all technical content is based on the sources listed in the bibliography.